\documentclass[12pt,a4paper]{article}
\usepackage{epsfig}
\usepackage{graphics}

\newcommand{\NIMA} {Nucl. Instr. and Meth. A}

\usepackage{amssymb}
\usepackage{epsfig}

\newcommand{\beq}[1]{\begin{equation}{\label{#1}}}
\newcommand{\eeq}[0]{\end{equation}}

\newcommand{\Fig}[1]{Figure~\ref{#1}}

\begin{document}
\title{{\bf Tagging heavy flavours with \\ boosted decision trees}}

\author{J. Bastos\footnote{bastos@lipc.fis.uc.pt} \\
{\small {\sl LIP-Coimbra, Universidade de Coimbra, P-3004-516 Coimbra}}}
\date{ }
\maketitle

\begin{abstract}
This paper evaluates the performance of boosted decision trees for tagging b-jets.
It is shown, using a Monte Carlo simulation of $WH \to l\nu q\bar{q}$ events
that boosted decision trees outperform feed-forward neural networks.
The results show that for a b-tagging efficiency of 60\% the light jet rejection given
by boosted decision trees is about 35\% higher than that given by neural networks. 
 
\end{abstract}


\section{Introduction}
\label{introduction}

Precision measurements in the top quark sector, and searches for the Higgs boson and physics beyond
the Standard Model, critically depend on the good identification (``tagging") of jets produced by b quarks.
Tagging techniques exploit specific properties of B-hadrons to differentiate them from the large background
of jets produced by light quarks and gluons.
The long lifetime of B-hadrons results in displaced vertices formed by tracks from their decays.
Physical observables associated to these vertices constitute the input for secondary vertex tagging.
Also, tracks from B- and D-hadron decays typically have large impact parameters, which are frequently used
to construct discriminating variables.
In a different approach, soft-lepton tagging searches for low transverse momentum leptons inside jets,
originating from semileptonic decays of B- and D-hadrons.
The tagging performance is substantially improved when individual taggers are combined to give a single
jet classifier.
In high energy physics, the feed-forward neural network is one of the most popular methods of combining
several discriminating variables into one classifier and have been extensively applied to b-tagging.

In this paper, the capability of an alternative classification technique, the boosted
decision trees, for tagging b-jets is evaluated.
Using a sample of $WH \to l\nu q\bar{q}$ Monte Carlo events, the performance of boosted
decision trees and feed-forward neural networks is compared.
Boosted decision trees is a learning technique recently introduced in high energy physics
for data analysis in the MiniBooNE experiment \cite{roe}. It was found that particle identification with
boosted decision trees has better performance than that with neural networks in a Monte Carlo simulation
of MiniBooNE data. This insight motivated the studies reported here, which indicate that boosted decision
trees is also a promising technique for tagging b-jets.

In the next section, a brief description of the boosted decision trees algorithm is given.
The Monte Carlo simulation used in this analysis is explained in Section \ref{monte_carlo_simulation}.
Section \ref{discriminant_variables} describes the discriminant variables which feed the tagging algorithms.
The tagging performances of boosted decision trees and neural networks are compared in Section \ref{results}.
Finally, conclusions are given in Section~\ref{conclusions}.

\section{Boosted decision trees}
\label{boosted_decision_trees}

\begin{figure}[htb]
\centering
\epsfig{file=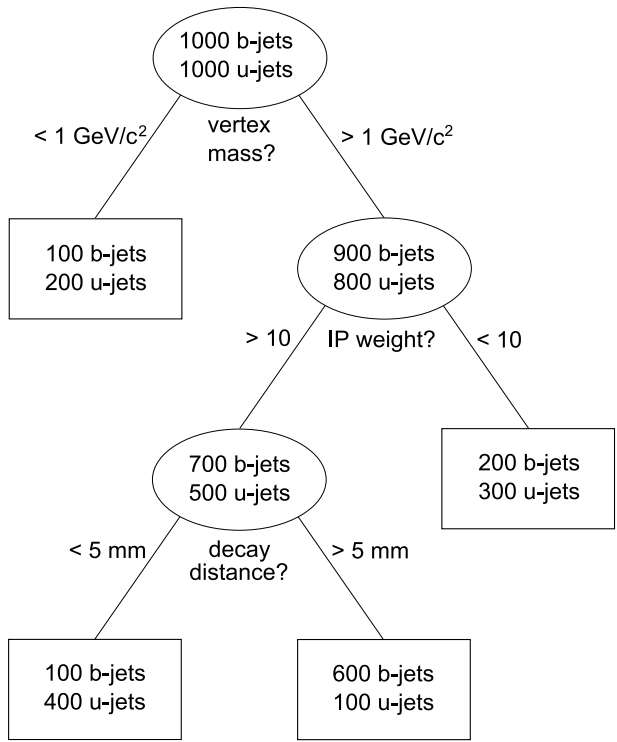, width=8cm}
\caption{Example of a decision tree.}
\label{BDT}
\end{figure}

The boosted decision trees algorithm implemented in this analysis starts with a parent node containing a
training set of b-jet and u-jet patterns.
All jets in the first tree iteration are given the same weight $w^{(0)}$, such that the sum of weights
equals 1.
Then, the algorithm loops over all binary splits in order to find the discriminating variable and
corresponding separation value that optimizes a given figure of merit.
For instance, in \Fig{BDT} the optimal figure of merit is obtained when the jets are divided
between those that have a secondary vertex mass greater than 1 GeV/c$^2$ and those that do not.
This procedure is then repeated for the new daughter nodes until a stopping criterion is satisfied.

A node is called ``signal node" if the sum of the weights of b-jets is greater than the sum of
the weights of u-jets. Otherwise, it is called ``background node".
A b-jet (u-jet) is correctly classified if it lands on a signal (background) node.
If $p$ designates the fraction of correctly classified jets in a node, its Gini index is defined to be
$Q(p) = -2p(1-p)$.
The optimal discriminating variable and separation value are the ones which maximize the figure of merit
\beq{qsplit}
   Q_{split} = \frac{w_LQ(p_L) + w_RQ(p_R)}{w_L+w_R}\,,
\eeq
where $w_L$ and $w_R$ are the sum of the jet weights in the left and right daughter nodes,
respectively, and $Q(p_L)$ and $Q(p_R)$ are the Gini indices of the left and right daughter nodes.
A node is not split if the optimal $Q_{split}$ is smaller than its own $Q(p)$, or, alternatively, if it contains
less events than a prespecified limit. Unsplit nodes are called ``leafs", which are depicted as
rectangles in \Fig{BDT}.

After the $k$th tree is built, the jet weights are updated. There are several methods to accomplish this.
Here, we will consider the AdaBoost algorithm~\cite{adaboost}.
First, the total misclassification error $\varepsilon_k$ of the tree is calculated:
\beq{epsilon}
   \varepsilon_k = \frac{\sum_{i=1}^{N_{jets}}w_i^{(k)} I_i^{(k)}}{\sum_{i=1}^{N_{jets}}w_i^{(k)}}\,,
\eeq
where $i$ loops over all jets in the training sample and $I_i^{(k)}$ is an indicator function which is equal
to 1 if the $i$th jet was misclassified or equal to 0 if the $i$th jet was correctly classified.
Then, the weights of misclassified jets are increased ({\it boosted})
\beq{w1}
w_i^{(k+1)} = \frac{w_i^{(k)}}{2\varepsilon_k}\,,
\eeq
while the weights of correctly classified jets are decreased
\beq{w2}
w_i^{(k+1)} = \frac{w_i^{(k)}}{2(1-\varepsilon_k)}\,.
\eeq
Finally, the tree $k+1$ is constructed using the new weights $w^{(k+1)}$.

After $M$ trees are trained their performance can be evaluated with a testing sample of
jets. The final score of jet $i$ is a weighted sum of the scores over the individual trees
\beq{score}
F_i = \sum_{k=1}^{M}\beta_kf_i^{(k)}\,,
\eeq
where
\beq{beta}
\beta_k = \log\left(\frac{1 - \varepsilon_k}{\varepsilon_k}\right)\,,
\eeq
and $f_i^{(k)} = 1 (-1)$ if the $k$th tree makes the jet land on a signal (background) leaf.
Therefore, b-jets will have large positive scores, while u-jets will have large negative scores.
Trees with lower misclassification errors $\varepsilon_k$ are given more weight when the jet score
is calculated.
Further details of the AdaBoost algorithm can be found in \cite{spr}.


\section{Monte Carlo simulation}
\label{monte_carlo_simulation}

The studies described in this paper were done with events generated with PYTHIA 6.319 \cite{pythia}.
We considered the environment of the LHC collider, in which $pp$ interactions with a center-of-mass energy of
14 TeV are produced.
One of the benchmark channels for b tagging studies at the LHC is the associated $WH$ production.
We generated $WH$ events with $m_H$ = 120 GeV/c$^2$, the $W$ boson decaying semileptonically
$W\to l\nu$ and the Higgs boson decaying to quark pairs $H\to q\bar{q}$.
Initial and final state radiation and multiple interactions were included in the simulation.

Tracks are parametrized by the following set of 5 parameters: $d_0$, $z_0$, $\phi$, $\cot\theta$ and $1/p_T$.
The transverse impact parameter $d_0$ is the distance of closest approach of the track to the primary vertex
in the plane perpendicular to the beam-line. The longitudinal impact parameter $z_0$ is the component along
the beam-line of the distance of closest approach. The parameters $\phi$ and $\theta$ are the azimuthal
and polar angles of the track, respectively, and $1/p_T$ is the inverse of the particle transverse momentum.

In order to simulate measurement errors, these parameters were smeared with Gaussian resolution functions.
The transverse and longitudinal impact parameters were smeared with standard deviations
$\sigma_{d_0} = 10$~$\mu$m and $\sigma_{z_0} = 100$~$\mu$m,
the angle $\phi$ with $\sigma_{\phi} = 0.10$~mrad, $\cot\theta$ with $\sigma_{\cot\theta} = 0.001$ and the
inverse of the transverse momentum with $\sigma_{1/p_T}= 0.001$ GeV$^{-1}$. The primary vertex positions were smeared
with Gaussian resolution functions with $\sigma_x = \sigma_y = 50$~$\mu$m and $\sigma_z = 100$~$\mu$m.
A jet is formed by all stable particles inside a cone $\Delta R  = \sqrt{(\Delta\phi)^2 + (\Delta\eta)^2} < 0.4$
around its axis, where $\eta = -\log\left(\tan(\theta / 2)\right)$ is the track pseudorapidity.


\section{Discriminant variables}
\label{discriminant_variables}

The physical observables used for discrimination between b-jets and light jets are taken from well known
``spatial" b-tagging algorithms. Physical observables from tagging techniques based on soft leptons are not
considered in this analysis.
Only jets with $p_T > 15$ GeV/c and $|\eta| < 2.5$ are considered taggable.

\subsection{Impact parameter tag}
\label{impact_parameter_tag}

\begin{figure}[htb]
\centering
\epsfig{file=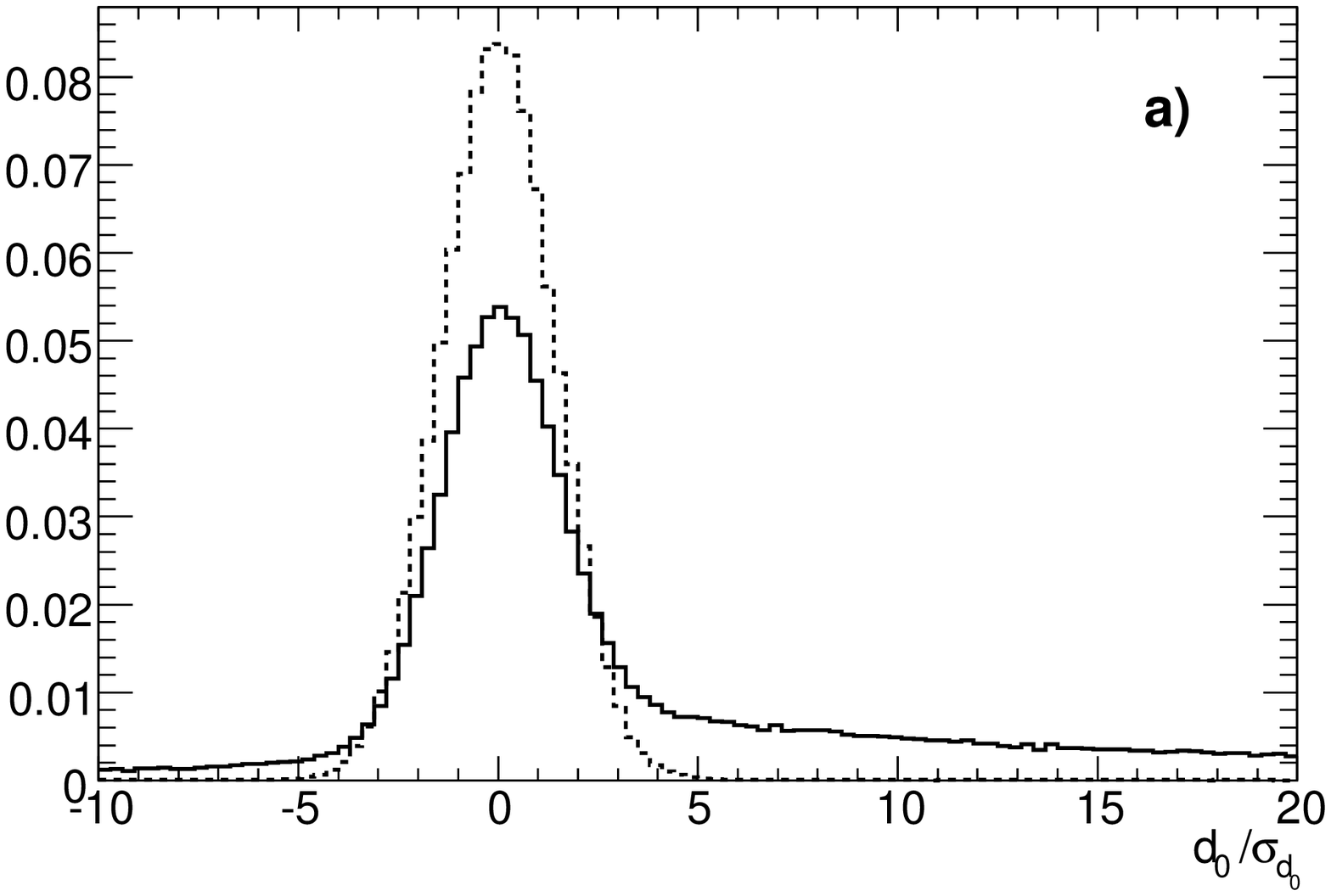, width=8cm} \\
\epsfig{file=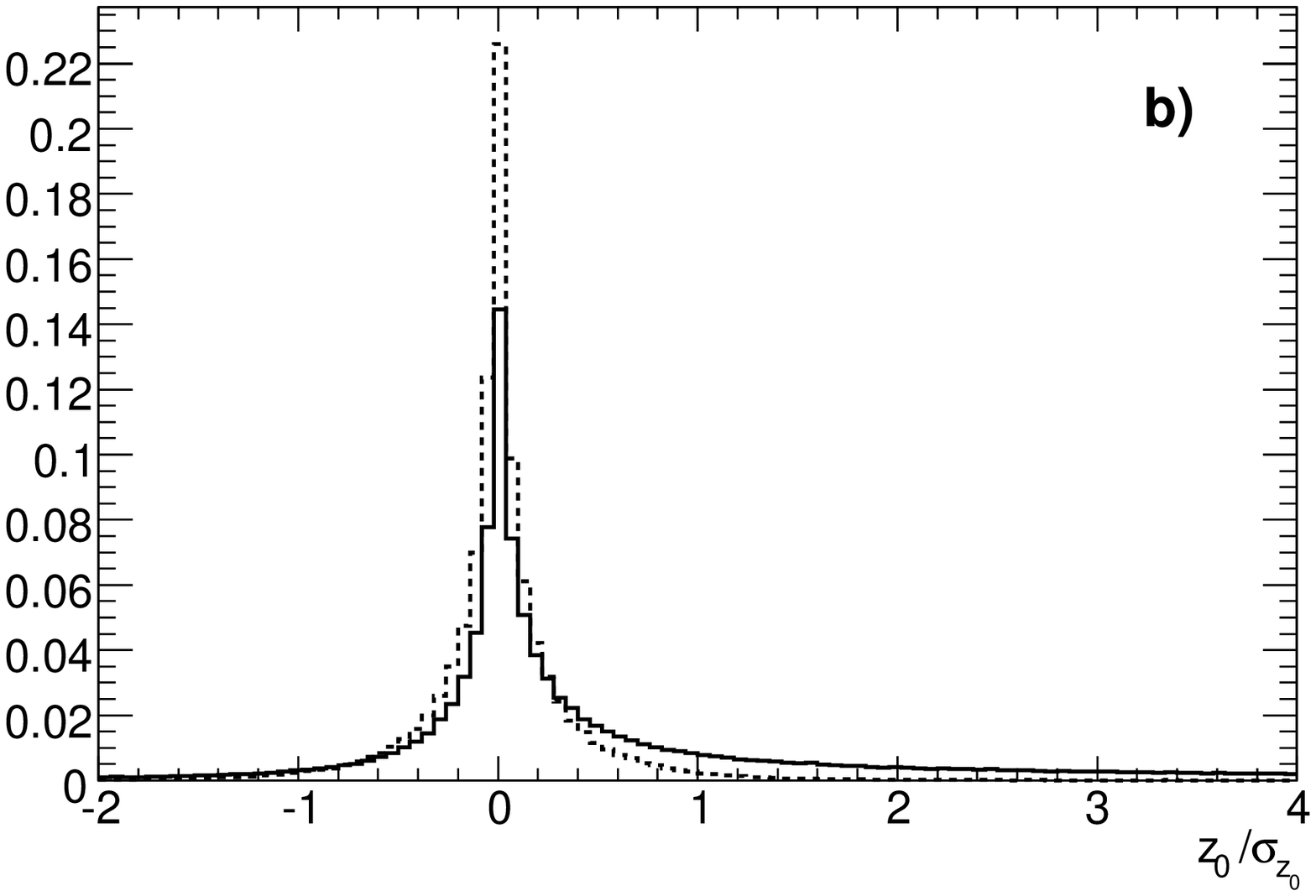, width=8cm}
\caption{(a) transverse and (b) longitudinal impact parameter significances for tracks in b-jets (solid line)
and u-jets (dashed line).}
\label{ipsig}
\end{figure}

Due to the long decay distances traveled by B-hadrons, tracks from b-jets have on average larger impact parameters
than tracks from light jets, since sizeable impact parameters in light jets are exclusively due to measurement errors.
Therefore, the impact parameter of jet tracks can be used to build a useful variable for discrimination between
b-jets and light jets.
\Fig{ipsig} shows the distributions of (a) signed transverse impact parameter significances $S_{d_0} = d_0 / \sigma_{d_0}$
and (b) signed longitudinal impact parameter significances $S_{z_0} = z_0 / \sigma_{z_0}$ of tracks in b-jets (solid line)
and u-jets (dashed line).
A positive (negative) sign is assigned to the impact parameter if the track intersects the jet axis in front (behind)
of the primary vertex.
These distributions give likelihood functions $b(S)$ and $u(S)$ for a track to belong to a
b-jet or a u-jet, respectively.
A jet weight is defined as the sum of the log-likelihood ratio over all tracks in the jet:
\beq{weight}
   w_{jet} = \sum_{i\in jet} \ln \left(\frac{b(S_i)}{u(S_i)}\right)\,.
\eeq

\begin{figure}[htb]
\centering
\epsfig{file=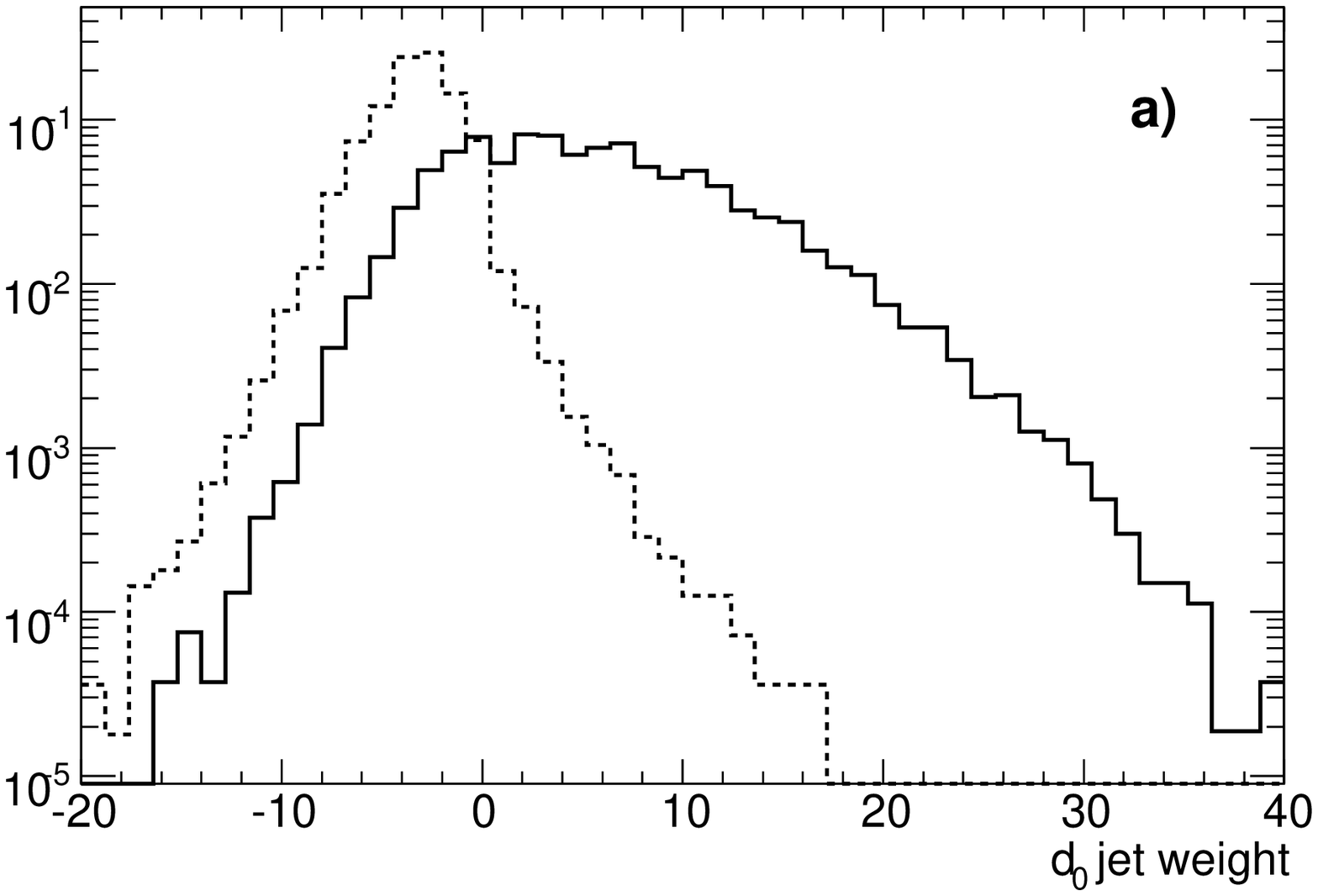, width=8cm} \\
\epsfig{file=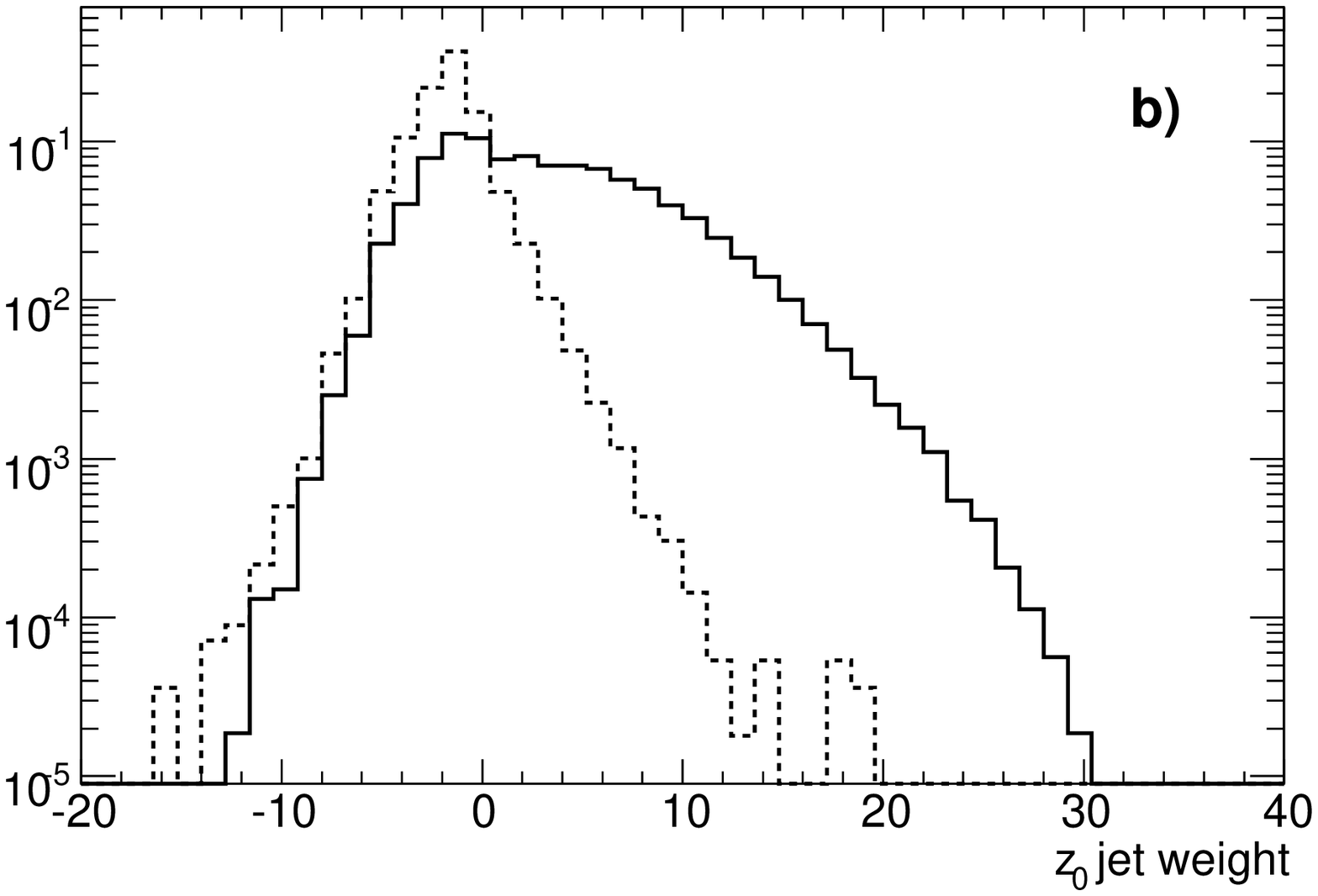, width=8cm}
\caption{Jet weight distributions given by the transverse impact parameter (a) and longitudinal impact parameter (b).
The solid (dashed) line corresponds to b-jets (u-jets).}
\label{weights}
\end{figure}

In \Fig{weights} it is shown the distribution of jet weights for u and b quarks.
Because the transverse impact parameter has better resolution, it yields greater discrimination power.
A given efficiency for selecting b-jets is obtained by selecting jets with weights above some
threshold level. Obviously, for moderate or high selection efficiencies there will always be some contamination
with light jets.

\subsection{Secondary vertex tag}
\label{secondary_vertex_tag}

\begin{figure}
\centering
\epsfig{file=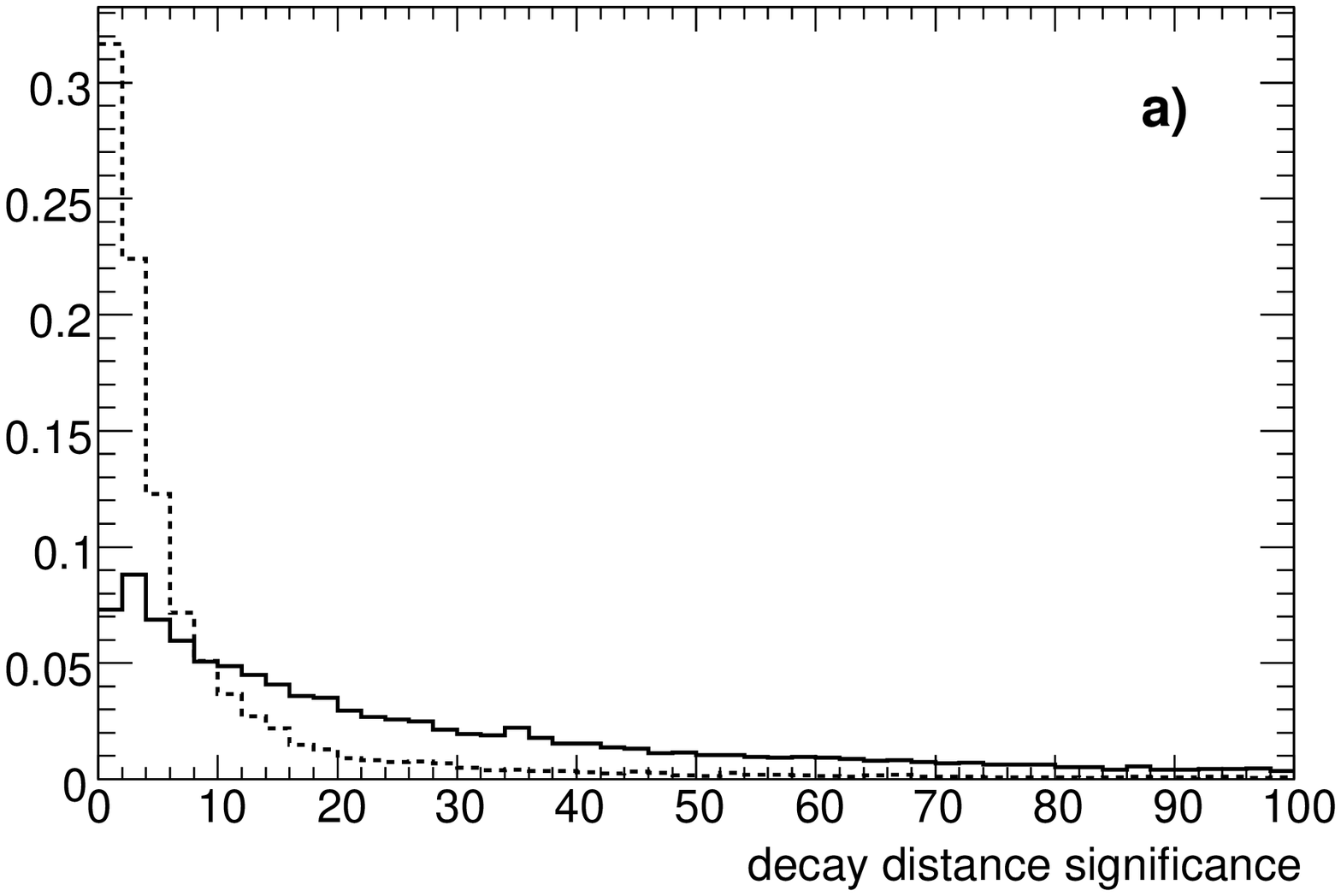,      width=7cm} \\
\epsfig{file=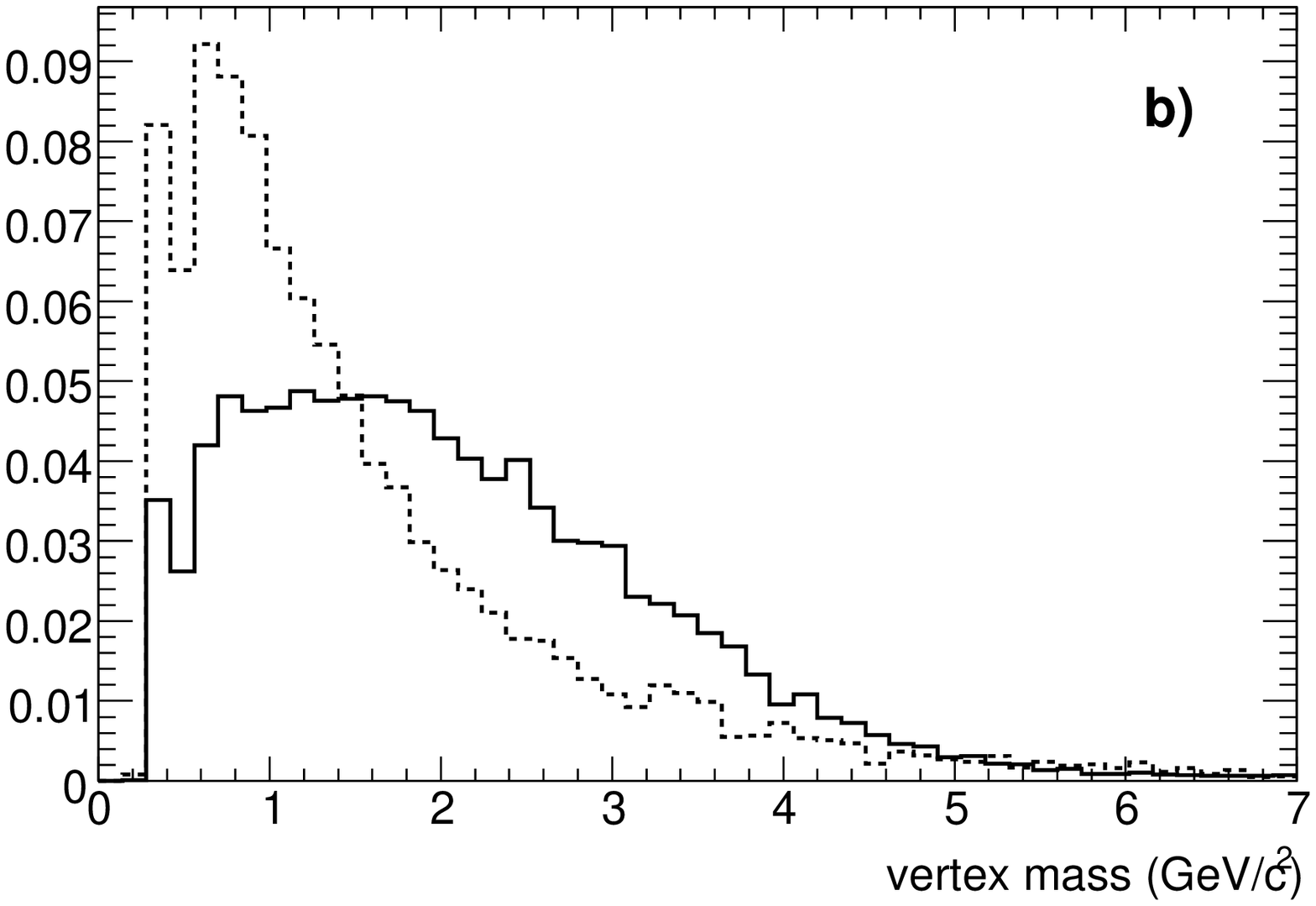, width=7cm} \\
\epsfig{file=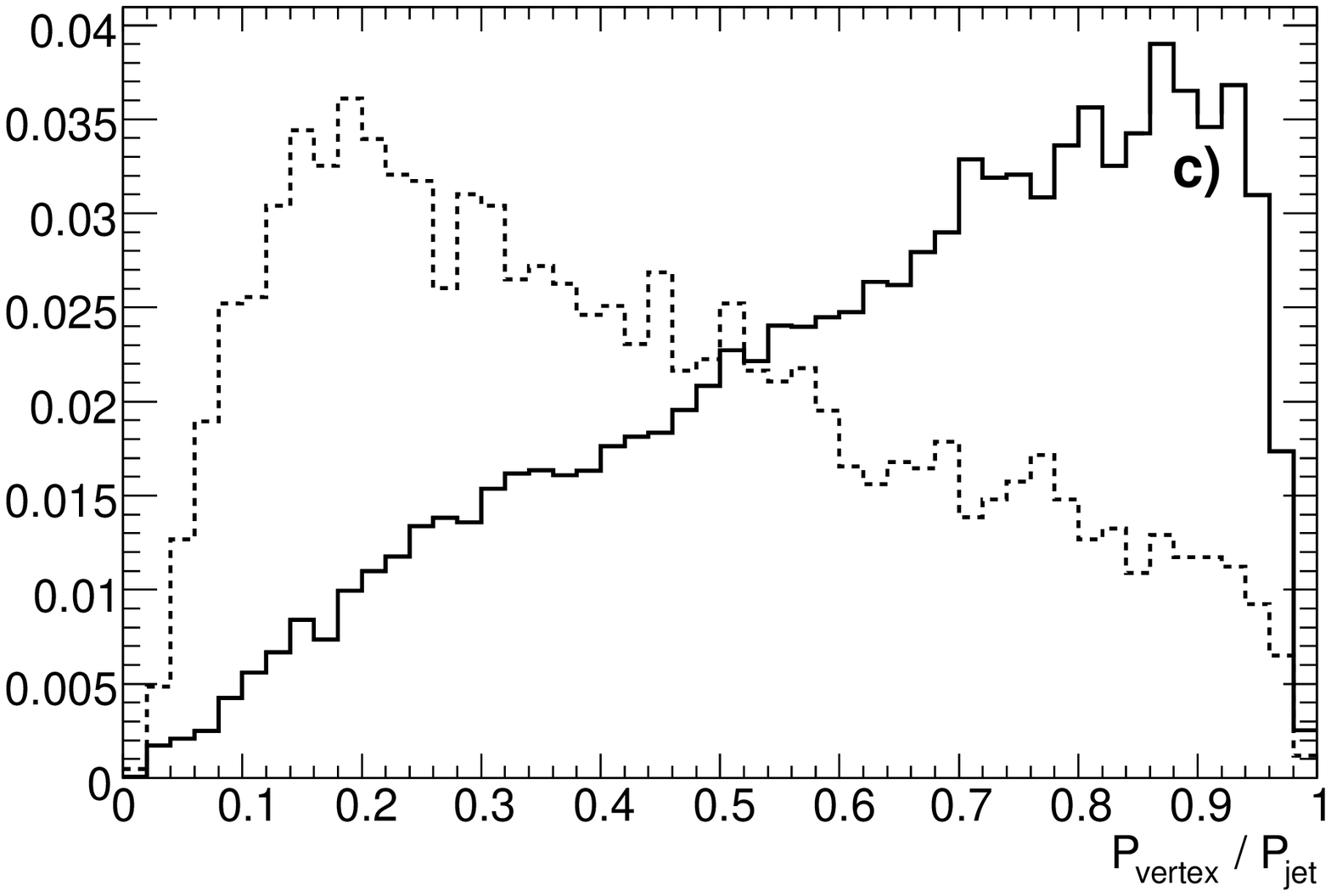,   width=7cm}
\caption{(a) Decay distance significance of the secondary vertex. (b) Invariant mass of tracks associated to the secondary vertex.
(c) Fraction of jet momentum in the secondary vertex. The solid (dashed) line corresponds to b-jets (u-jets).}
\label{vertex}
\end{figure}

An alternative approach for building b tagging discriminating variables consists in reconstructing displaced
secondary vertex from B- and D-hadron decays inside the jet. Secondary vertices were reconstructed with
Billoir and Qian's fast vertex fitting algorithm \cite{billoir_qian}.
For purposes of secondary vertex b-tagging the exact topology of the secondary vertex is irrelevant and,
therefore, an inclusive vertex search is performed.
All jet tracks with large transverse impact parameter significance participate in the vertex fit
and vertices compatible with $V^0$ decays are rejected.
\Fig{vertex}(a) shows the decay distance significance for b-jets and u-jets for good quality vertices.
Besides the decay distance significance, other variables associated to the secondary vertex may have discrimination
power, such as the vertex mass (\Fig{vertex}(b)) and the ratio between the absolute momentum sum of tracks in the
secondary vertex and that of all tracks in the jet (\Fig{vertex}(c)).

\subsection{One-prong tag}
\label{one_prong_tag}

\begin{figure}[htb]
\centering
\epsfig{file=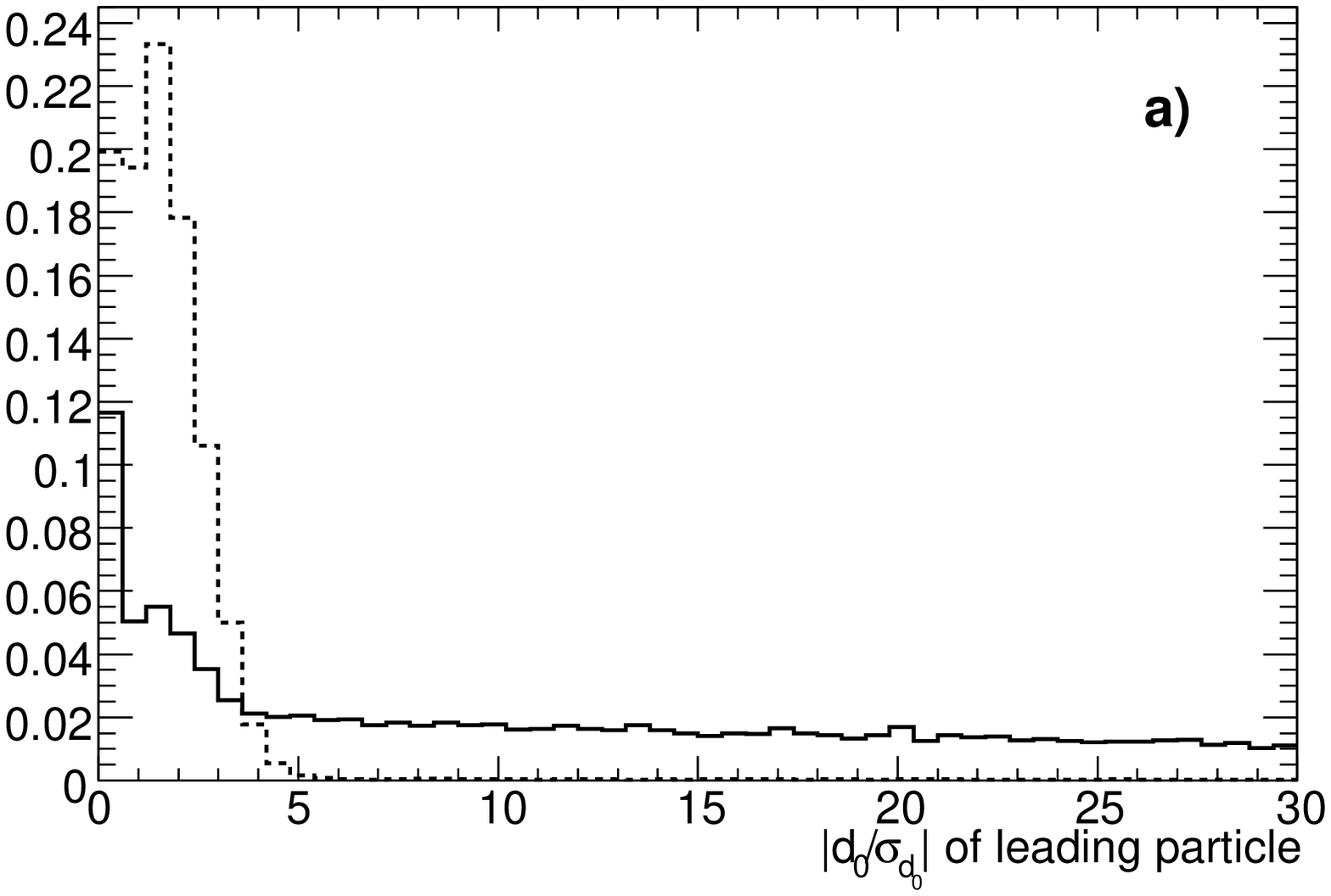, width=8cm} \\
\epsfig{file=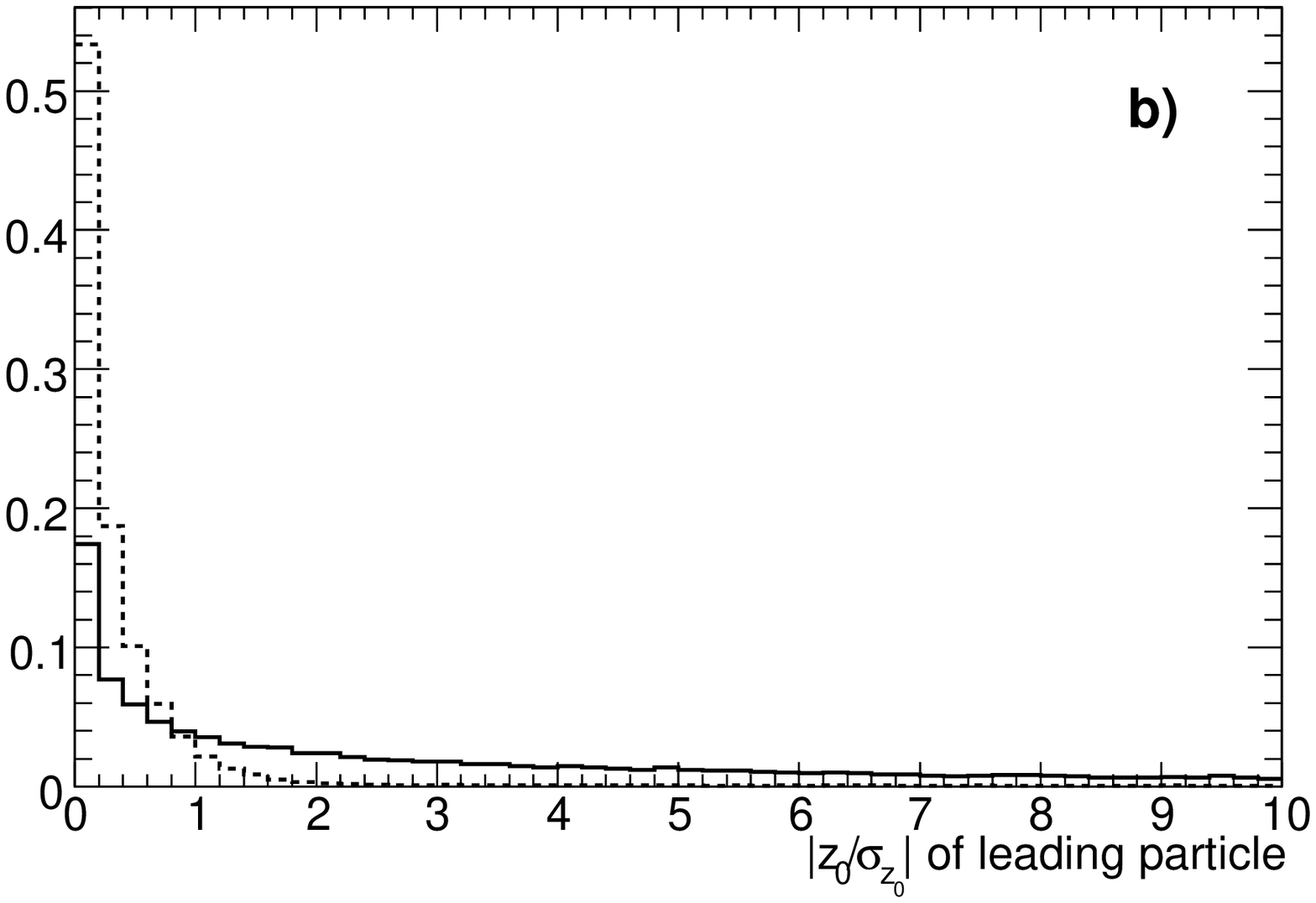, width=8cm}
\caption{Maximal (a) transverse and (b) longitudinal impact parameter significances in jets.
The solid (dashed) line corresponds to b-jets (u-jets).}
\label{maxip}
\end{figure}

For one-prong decays of B- and D-hadrons the secondary vertex fit fails.
In this situation, though, some information can still be extracted from tracks in the jet.
For instance, the maximal transverse and longitudinal impact parameters of jet tracks clearly have
discrimination power, as can be observed in \Fig{maxip}.

\section{Results}
\label{results}

\begin{figure}[htb]
\centering
\epsfig{file=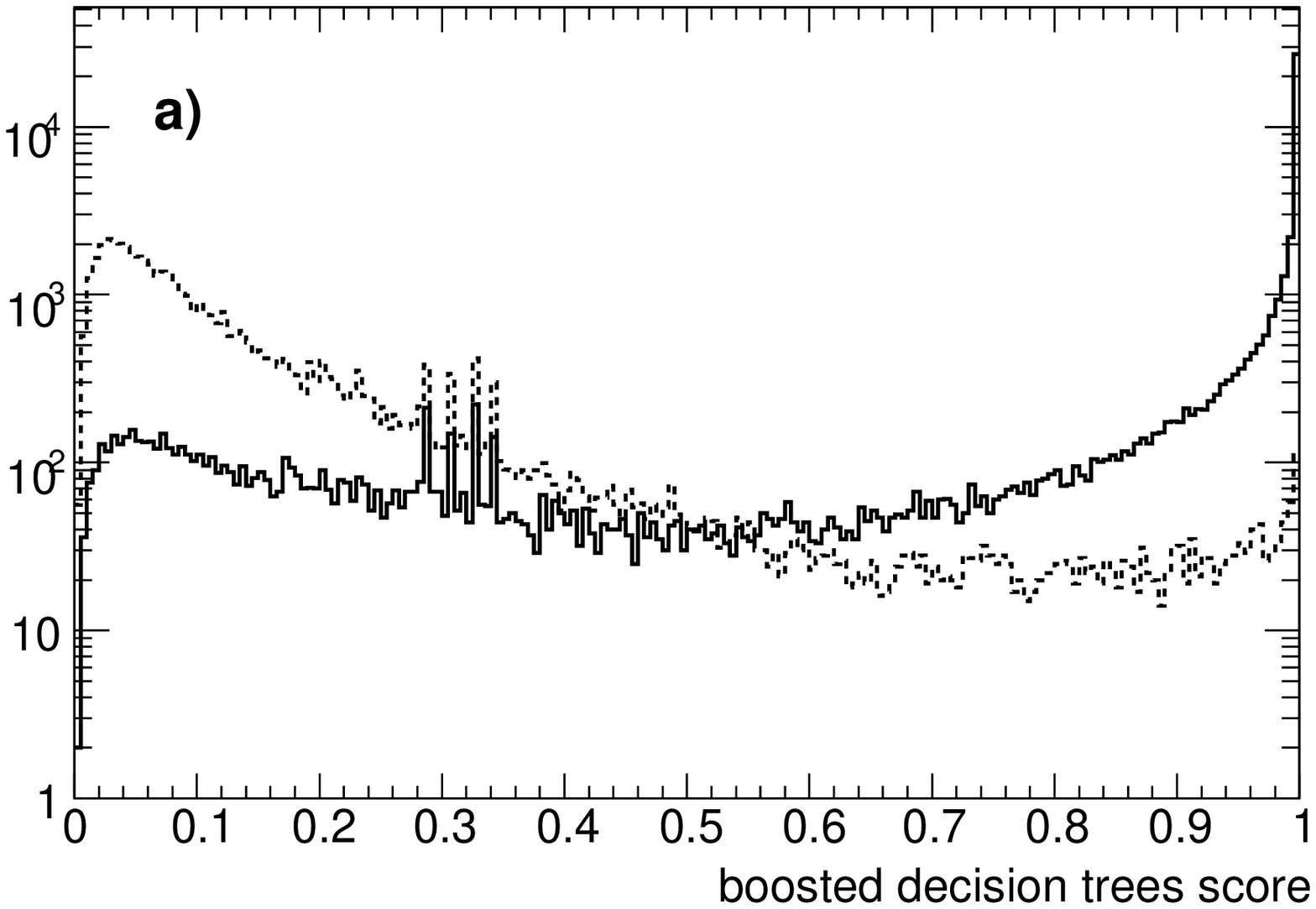, width=8cm} \\
\epsfig{file=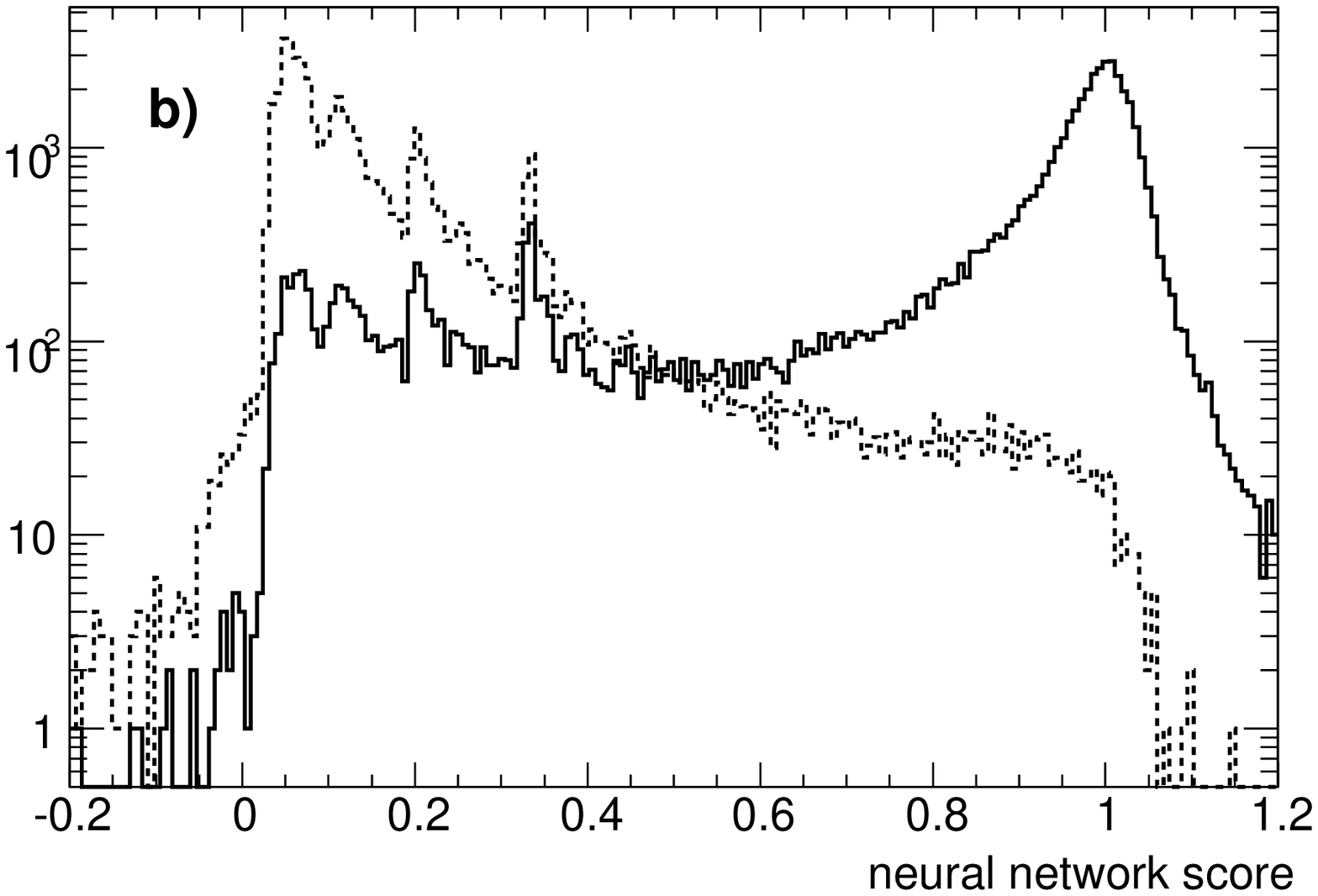,  width=8cm}
\caption{Jet scores given by (a) boosted decision trees and (b) a neural network, for b-jets (solid line) and
u-jets (dashed line).}
\label{scores}
\end{figure}

Boosted decision trees were implemented using the StatPatternRecognition package \cite{spr}.
The trees were fed with the 7 discriminant variables mentioned in the previous section
and were trained with 50000 b-jet patterns and 50000 u-jet patterns.
An unbiased evaluation of the boosted decision trees performance is obtained using a distinct sample of
b-jets and u-jets patterns (test sample).
The best results were obtained with a minimum number of jets per leaf of about 7000.
The performance becomes better with increasing number of trees, but no significant improvement was observed
after several hundreds of tree iterations.
\Fig{scores}(a) shows the jet scores, normalized to be within the interval $\left[0,1\right]$,
for the test sample of b-jets (solid line) and u-jets (dashed line).

In order to compare the performance of boosted decision trees with the neural network approach, a
feed-forward neural network was implemented using the Multi-Layer Perceptron class \cite{mlproot}
provided by the data analysis framework ROOT \cite{root}.
The architecture of the network consisted of 7 nodes in the input layer (corresponding to the 7 discriminant variables
mentioned above), 8 nodes in a single hidden layer and 1 node in the output layer.
The network was trained with the Broyden-Fletcher-Goldfarb-Shanno learning method with a learning rate parameter
$\eta = 0.1$.
The training set consisted of 100000 jet patterns, of which 50000 were b-jets and 50000 were u-jets.
Since the magnitude of the discriminant variables differ considerably, which may affect the performance of the
neural network, all input variables were normalized.
The number of epochs (training cycles) was 1000.
Care was taken to prevent overtraining the network by monitoring the evolution of the learning curve.
\Fig{scores}(b) shows the jet scores given by the neural network for a test sample of b-jet (solid line)
and u-jet (dashed line) patterns.

\begin{figure}[htb]
\centering
\epsfig{file=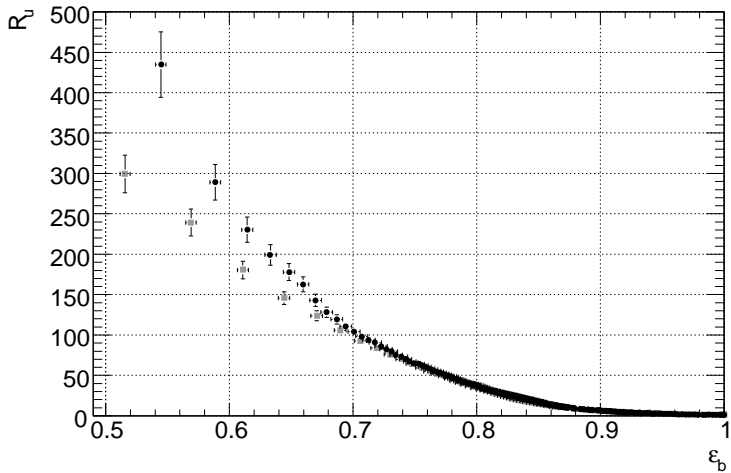, width=10cm}
\caption{Light jet rejection as a function of b-jet efficiency given by boosted decision trees (black circles)
and a feed-forward neural network (gray squares).}
\label{comparison}
\end{figure}

Jets with a score above some specified threshold value are tagged as b-jets. The threshold value is contingent on
the desired efficiency for tagging b-jets $\varepsilon_b = N_b^{tag} / N_b$, where $N_b$ is the number of b-jets in
the data and $N_b^{tag}$ is the number of tagged b-jets, or, alternatively, on the tolerated level of contamination
by light jets.
\Fig{comparison} shows the light jet rejection, $R_u = \varepsilon_u^{-1}$ as a function of the b-tagging
efficiency $\varepsilon_b$, for the test sample given by boosted decision trees (black circles) and the
feed-forward neural network (gray squares).
For high b-tagging efficiencies there is no significant improvement of the performance of boosted decision trees
relative to the neural network.
However, for moderate b-tagging efficiencies, boosted decision trees clearly outperform neural networks.
For a b-tagging efficiency of 60\%, the light jet rejection given by boosted decision trees is about 35\% higher than
that given by neural networks.

\section{Conclusions}
\label{conclusions}

The studies presented in this paper indicate that boosted decision trees outperform neural networks for
tagging b-jets, using a Monte Carlo simulation of $WH \to l\nu q\bar{q}$ events, and sensible physical
observables as discriminating variables.
For a b-tagging efficiency of 60\%, the light jet rejection given by boosted decision trees is about 35\% higher than
that given by the neural network approach.
Although encouraging, these results should be complemented with studies performed with a full simulation
in which detector inefficiencies are considered.
Also, the performance of both techniques may differ if other physics channels are considered, since it may be affected
by jet overlaps and gluon splitting into $b\bar{b}$ pairs.

\section*{Acknowledgments}
I would like to thank J. Carvalho and A. Onofre for many valuable remarks.
This work was supported by grant SFRH/BPD/20616/2004 of Funda\c c\~ao para a Ci\^encia e Tecnologia.

\end{document}